\documentclass[aps,pre,twocolumn,showpacs,superscriptaddress]{revtex4}
\usepackage{graphicx}
\begin{document}

\title{Dynamic critical behavior of the \textsl{XY} model in
  small-world networks} 
\author{Kateryna Medvedyeva}
\email{medv@tp.umu.se}
\affiliation{Department of Physics, Ume{\aa} University, 901 87
  Ume{\aa}, Sweden} 
\author{Petter Holme}
\affiliation{Department of Physics, Ume{\aa} University, 901 87
  Ume{\aa}, Sweden} 
\author{Petter Minnhagen}
\affiliation {NORDITA, Blegdamsvej 17, DK-2100, Copenhagen, Denmark}
\affiliation{Department of Physics, Ume{\aa} University, 901 87
  Ume{\aa}, Sweden}

\author{Beom Jun \surname{Kim} }
\affiliation{Department of Molecular Science
  and Technology, Ajou University, Suwon 442-749, Korea}

\begin{abstract}
The critical behavior of the \textsl{XY} model on small-world network
is investigated by means of dynamic Monte Carlo simulations. We use
the short-time relaxation scheme, i.e., the critical behavior is
studied from the nonequilibrium relaxation to equilibrium. 
Static and dynamic critical exponents are extracted 
through the use of the dynamic finite-size scaling analysis.
It is concluded that the dynamic universality class at the 
transition is of the mean-field nature. We also confirm numerically 
that the value of dynamic critical exponent is independent
of the rewiring probability $P$ for $P \gtrsim 0.03$.
\end{abstract}
\pacs{89.75.Hc, 64.60.Fr, 74.20.-z}

\maketitle

\begin{figure}
\resizebox{7.5cm}{!}{\includegraphics{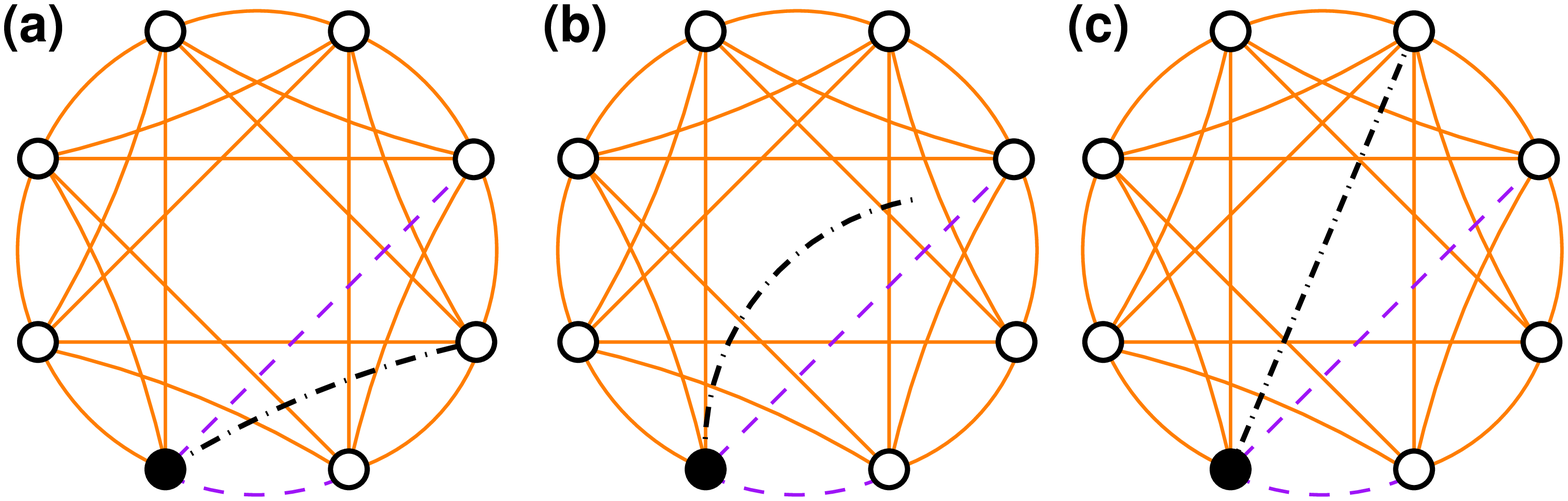}}
\caption{The construction of a Watts-Strogatz model network. One
  starts from a regular one-dimensional lattice (a). For every vertex
  (we consider the black vertex specifically) one goes through the
  edges on one side (for the black vertex these are dashed). Then,
  with probability $P$ one detaches the other end (b) and re-attach it
  (c) with the condition that no loops (edges starting and ending at
  the same vertex) or multiple edges must be formed.}
\label{fig:ws_col}
\end{figure}

\section{Introduction}
In recent years, there has been a surge of activity in the field of
complex networks among statistical and interdisciplinary
physicists~\cite{rev}. Quite naturally,
various spin models of statistical mechanics have been
studied on an underlying complex network~\cite{barr,statswn,spin}. 
These studies serve a twofold purposes: 
Firstly, they aid studies of the static network
structure. In many real-world situations, the network structure
is an underlying infrastructure for a dynamical system, and
non-trivial effects can emerge from the interplay between the
dynamical system and the network~\cite{traffic}. 
Secondly, such studies of spins systems on complex networks can illuminate the
properties of the spin-model itself in certain extreme situations. For
example, both the Ising and \textsl{XY} models can display a
critical behavior similar to high dimensional regular lattices with a
very low density of couplings (or edges in the network) between
spins~\cite{barr,statswn}.

One of the most central complex network models is the Watts and
Strogatz (WS) model of small-world networks~\cite{WS}. Briefly,
this model is controlled by a parameter $P$ (the ``rewiring probability''),
and by tuning $P$ from 0 to 1 one goes from regular to random
networks. The interesting region is that of intermediate $P$ where the
network is clustered (has a high density of short circuits, or more
specifically, triangles) and a logarithmically increasing average
path-length (the path length of a pair of vertices is the smallest
number of intervening edges). In the \textsl{XY} model, 
each vertex is associated with a two-dimensional spin-angle. 
The \textsl{XY} model has mostly been used to study phase transitions in 
superconductors and superfluids, while it was also applied to e.g., 
the formations of bird flocks~\cite{birds}. The static properties of 
the \textsl{XY} model in the WS network have been studied in \cite{statswn}, 
where critical exponents characteristic of a mean-field transition have been
found at any nonzero value of $P$. In the present paper, we study 
the \textit{dynamic critical behavior} of the \textsl{XY} model on the WS small-world
network with focus on the dynamic critical exponent.

\section{\textsl{XY} model on WS model network}
In the WS model for the small-world network~\cite{WS}, a regular
network is first constructed by arranging $N$ vertices 
in a one-dimensional circular topology and connecting each vertex
to $2k$ neighbors. Then one goes through each edge one at a time, 
and with the rewiring probability $P$ detaches the far-side of the edge and
reconnect it to a randomly chosen other vertex (with the restriction
that loops and multiple edges must not be formed). In this manner, a
small-world network with the size $N$ is constructed with the model
parameters $k$ and $P$. This procedure is illustrated
Fig.~\ref{fig:ws_col} The former parameter ($k$)  is not
believed to give any significant change of the network structure for
$k>1$, and thus we fix $k=3$ throughout the paper.

The \textsl{XY} model consists of planar spins interacting through the Hamiltonian
\begin{equation}\label{eq:hamiltonian}
  H = -{1\over 2} \sum_{i \neq j} J_{ij}\cos(\theta_i-\theta_j)~,
\end{equation}
where $\theta_i\in [-\pi,\pi)$ at vertex $i$ is the spin angle, corresponding
to the phase of the superconducting order parameter in the  Ginzburg-Landau
theory of superconductivity. The coupling matrix $J_{ij}$ is given by
\begin{eqnarray}\label{eq:coupl}
J_{ij} = J_{ji}=\left\{\begin{array}{ll}
 J, & \mbox{\vspace{1cm} if $(i,j)$ is an edge,}\\
    0, & \mbox{\vspace{1cm} otherwise.}
    \end{array}
    \right.   
\end{eqnarray}
For example, in the \textsl{XY} model on a
two-dimensional square lattice where only nearest vertices interact, we
have $J_{ij}=0$ except when $i$ and $j$ are nearest neighbors. For convenience,
we measure the temperature $T$ in units of $J/k_B$.

\section{Short-time relaxation method and scaling analysis} \label{sec:short}
To investigate the dynamic critical behavior of the \textsl{XY} model
on the WS network we use the so-called "short-time relaxation method", which 
utilizes the relaxation behavior of the system towards equilibrium from
the a nonequilibrium initial state. By use of this method several
critical exponents have been successfully determined for the
Ising model~\cite{shorttime,str}, for unfrustrated and fully
frustrated Josephson Junction arrays~\cite{luo}, and for the
classical Heisenberg spin \textsl{XY} model~\cite{spinxy}. The major advantage
of the short-time relaxation method (compared to dynamical simulations
in equilibrium) is the running time saved from the avoidance of
equilibration. 

The Monte Carlo (MC) scheme is based on the
Hamiltonian~(\ref{eq:hamiltonian}) and the standard Metropolis local
update algorithm~\cite{metropolis,kim}. 
The key quantity we measure is~\cite{str,scaling1}
\begin{equation}\label{eq:psi}
   Q(t) = \Biggl [\biggl\langle \mbox{sgn}  \biggl(\sum_{i=1}^{N} \cos
   \theta_{i}(t)\biggr)\biggr\rangle \Biggr ]~,
 \end{equation}
where the time $t$ is measured in units of one MC sweep, 
$\langle\: \cdots\: \rangle$ is the average over different time
sequences from the same starting configuration, and average over
different network configurations, denoted by $[\: \cdots\: ]$, should
also be taken. Here the sign function $\mathrm{sgn}(x)$ measures the
sign ($\pm 1$) of $x$. The initial configuration is
chosen as $\theta_{i}(0)=0$, giving $Q(0)=1$, and $Q(t \rightarrow \infty) = 0$
since in equilibrium ($t \rightarrow \infty )$ there is no preferred angle direction.
We chose the trial angle
$\delta \theta = \pi/6$; the motivation is that it is sufficiently
small in order to obtain good convergence rate of the quantity we
measure while it is big enough to make simulations fast~\cite{kim}.

In order to obtain the dynamic critical exponent and detect
the phase transition, we use the finite-size scaling of the quantity $Q$.
Close to the critical temperature $T_c$ one expects that in a finite-sized system the
characteristic time $\tau$ scales as $\tau \sim N^{\bar z}$,
while the ratio of the correlation volume $\xi_V \sim
|T-T_c|^{-\bar \nu}$ to the system size $N$ gives the second argument
of the scaling function~\cite{note,statswn,scaling1,scaling2}:
\begin{equation} \label{eq:scaling}
Q(t,T,N) = F\biggl(t/N^{\bar z}, (T-T_{c})N^{1/\bar \nu}\biggr)~,
\end{equation}
where $F(x_1,x_2)$ is the scaling function with the property
$F(0,x_2)=1$. At $T_c$, where the second scaling variable vanishes,
the dynamic exponent $\bar z$ is easily determined from Eq.~(\ref{eq:scaling}) by
the requirement that the $Q(t)$ curves obtained for different sizes
of the networks collapse onto a single curve when plotted against the
scaling variable $tN^{-\bar z}$. It is also
possible to determine $T_c$ from Eq.~(\ref{eq:scaling}) by applying an
intersection method: Starting from the fully phase ordered
nonequilibrium state, $Q$ decays from $1$ to $0$ as time proceeds. 
For times $t$ where $0 < Q(N,T,t) < 1$ we can fix the
parameter $a=tN^{-\bar z}$ to a constant for given $N$ and $\bar z$. Then
$Q$ has only one scaling variable $(T-T_c)N^{1/\bar \nu}$ and can thus
be written as
\begin{equation}\label{eq:scaling1}
Q_a(T,N) = F\biggl (a, (T-T_{c})N^{1/\bar\nu}\biggr)~.
\end{equation}
If now we plot $Q$ with fixed $a$ as a function of $T$ for various
$N$, all curves should have a unique intersection point at
$T=T_c$. Finally, we can check the consistency by using the full
scaling form to collapse the data for different temperatures and
networks sizes onto a single scaling curve in the variable
$(T-T_c)N^{1/\bar \nu}$ at fixed $a=tN^{-\bar z}$. In addition, this
is a consistency check of the value of the static
exponent $\bar \nu$.

To discuss the finite-size scaling in more detail, the form
Eq.~(\ref{eq:scaling1}) is based on the
assumption that there is only two length scales in the system: the
network size $N$ (or the number of vertices in the network) and the
correlation volume $\xi_V$ diverging at $T_c$. However, it is
known that in the small-world network there is an additional
spatial length scale related with the distance between shortcut
endpoints, given by $\zeta =(kP)^{-1}$~\cite{newman}. Accordingly, in
the presence of the three competing scales ($N$, $\xi_V$, and $\zeta$),
the finite-size scaling function should take the form $\chi(t/N^{\bar
  z},\xi_V/N,\zeta/N)$~\cite{newman,ozana}. Here, we aim to use
sufficiently large systems with $N$ much larger than $\zeta$ (but, as
we will see, this is difficult for small $P$), where
$\chi(t/N^{\bar z},\xi_V/N,\zeta/N)$ may be approximated as
$\chi(t/N^{\bar z},\xi_V/N,0)$. This leads to the above mentioned
scaling forms~(\ref{eq:scaling}) and (\ref{eq:scaling1}) without $\zeta$.

\section{Simulation results}

\begin{figure}
\resizebox{!}{5.5cm}{\includegraphics{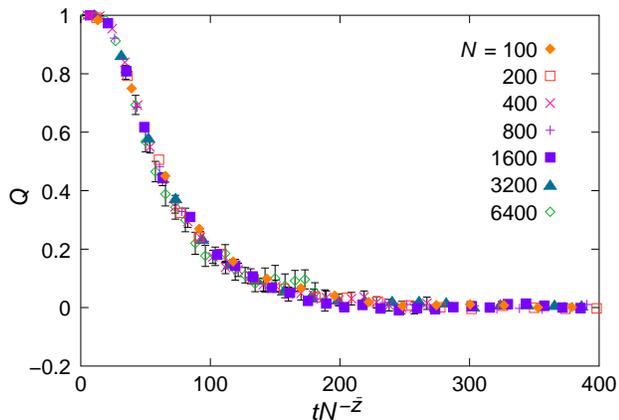}}
\caption{Short-time relaxation of $Q$ for $P=0.2$ at $T=T_c=2.23$.
  $Q$ is shown as a function of the scaling variable $tN^{-\bar
    z}$. $\bar z=0.52(1)$ is found at the best data collapse (see
  the Appendix).}
\label{fig:scaling}
\end{figure}

\begin{figure}
\resizebox{!}{5.5cm}{\includegraphics{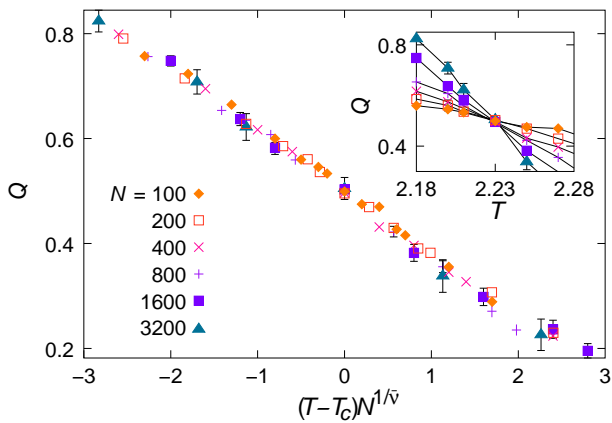}}
\caption{Finite-size scaling of the short-time relaxation of $Q$
  with $P=0.2$ and at $T=2.18$, $2.20$, $2.21$, $2.23$, $2.25$, and
  $2.27$. The inset an intersection plot with fixed $t/L^{\bar{z}}=a$
  and $(\bar{z},a)=(0.52,3.0)$; this is consistent with $T_c \approx
  2.23$, while the main part of the graph displays the full scaling of
  $Q=F(t/N^{\bar{z}},(T-T_c)L^{1/\bar{\nu}})$ with the
  mean-field value of $\bar\nu = 2$~\cite{statswn}.}
\label{fig:double}
\end{figure}

\begin{figure}
\resizebox{!}{5.5cm}{\includegraphics{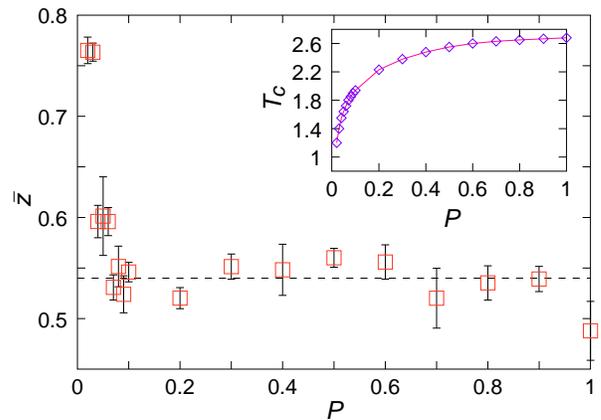}}
\caption{The dynamic critical exponent $\bar z$ as a function of the
  rewiring probability $P$. The inset showing $T_c$ as a function of
  $P$ is consistent with Fig.~4 in Ref.~\cite{statswn}. The dashed
  line is $\bar z=0.54$.}
\label{fig:z}
\end{figure}

We exemplify the critical behavior of $Q$ for WS model networks with
$P=0.2$. This value is quite representative for all $P$ values of our
simulations, but (as we will discuss later) small $P$ requires larger
system sizes, longer times series, and more averages. 
Figure~\ref{fig:scaling} shows the finite-size scaling of the
short-time relaxation given by Eq.~(\ref{eq:scaling}) which at $T_c$
turns into the simple form:
\begin{equation}\label{eq:scalingTc}
Q(t) = F(t/N^{\bar z},0) 
\end{equation}
with the only one scaling variable $t/N^{\bar z}$. 
In Fig.~\ref{fig:scaling} (as well as all other $P$ and $T$ values) we
have performed a sample average over $100$ independent runs
for $200$ different network realizations. Instead of leaving both ${\bar z}$
and $T_c$ as free parameters, we use $T_c$ obtained from
static MC simulations~\cite{statswn}. Figure~\ref{fig:scaling} displays
the best collapse onto a single curve in a broad range of the scaling
variable $tN^{-\bar z}$ with $\bar{z} = 0.52(1)$ where the number in
the parenthesis is the error in the last digit (how ${\bar z}$ is
obtained is described in detail in the Appendix)~\cite{note1}. Just as for static
quantities~\cite{statswn} the obtained $\bar{z}$ is consistent with
higher dimensional regular lattices ($d\geq 4$ to be precise), 
where $\bar z =0.5$ is expected~\cite{note}.

However, the above method presumes \textit{a priori} knowledge of
$T_c$. To check out the consistency of determination of $T_c$ one can
use an intersection method described above in Sec.~\ref{sec:short}. In the inset of
Fig.~\ref{fig:double} we display $Q$ as a function of $T$ for
different network sizes $N$ with a fixed value of $a=tN^{-\bar{z}}$ in
the first argument of the scaling form in Eq.~(\ref{eq:scaling}).
We find a unique crossing point at $T = T_c=2.23$ and
$\bar{z}=0.52$. In some cases (typically for small $P$ values) the
$T_c$ has to be slightly altered (from the values of
Ref.~\cite{statswn}) to get both the collapse and intersection plots
of Fig.~\ref{fig:double} correct. We
then use $\bar z$ and $T_c$ estimated as above to make the full
scaling plot for $Q$ as displayed in the main part of
Fig.~\ref{fig:double}. A very smooth collapse here is obtained with
$\bar{\nu}=2.0$ which is again consistent with Ref.~\cite{statswn}.

The procedure described above for $P=0.2$ is then repeated for various
values of $P$ to obtain Fig.~\ref{fig:z}. As one can see, except for $P
\lesssim 0.03$, $\bar{z} = 0.54(3)$~\cite{note:z} throughout the
broad range of $P$. We
believe that the nature of the transition (and hence $\bar{z}$) is
independent of $P$ for all $P>0$. The larger values of $\bar z$ for
small $P$ is a result of a failure of the assumption that we can
neglect the length scale $\zeta$ since $N \gg \zeta (\sim 1/P)$ cannot
be valid for small $P$. The inset of Fig.~\ref{fig:z} displays the
dependence of critical temperature $T_c$ (obtained as discussed above)
upon $P$ and is consistent with what has been obtained from static
MC~\cite{statswn}.

\section{Summary}
In conclusion, we have studied the dynamic critical behavior of the
\textsl{XY} model on WS model networks by means of dynamic Monte Carlo
simulations. We have used the short-time relaxation method, based on
the relaxation from a nonequilibrium state, and determined the
critical temperature $T_c$, the dynamic critical exponent $\bar z$, as
well as the static correlation-volume exponent $\bar \nu$. The dynamic
critical exponent was determined to be $\bar z = 0.54(3)$ for the
networks with rewiring probability $P \gtrsim  0.03$, while the static
critical exponent was found to be $\bar \nu \approx 2.0$. We believe that
this result will hold for any $P>0$ but that the system size needed
to confirm this diverges as $P\rightarrow 0$. The exponent $\bar \nu$, as
well as two others,  critical exponents $\alpha$ and $\beta$ of the
specific heat and magnetization respectively have been obtained in
Ref.~\cite{statswn}. The obtained values  $\bar \nu=2$, $\beta=1/2$, and
$\alpha=0$, which also have been shown to be independent from the
value of $P$, establish the mean-field nature of the transition in
\textsl{XY} model on WS networks. The result of the present paper
support this picture and since the upper dimensionality for the
mean-field theory is $d=4$, one can conclude that the phase transition
in \textsl{XY} model on WS networks is in the same universality class
as a regular lattice of dimensionality $d\geq 4$.

An interesting observation is that for a regular hyper-cubic
lattice this behavior requires a number of edges larger than $8N$,
whereas for in our simulations we have much fewer ($3N$) edges; and
most probably $k=2$ (giving $2N$ edges) gives the same behavior. We
also note that there is no additional critical behavior \textit{induced} by the WS model
other than the transition from linear (``large-world'') to
logarithmic (small-world) behavior in diameter as $P$ becomes finite.

\section*{Acknowledgements}
This work was partly supported by the Swedish Research Council through
contract no.\ 2002-4135.
B.J.K.\ was  supported by the Korea Science and Engineering
Foundation through Grant No.\ R14-2002-062-01000-0.

\appendix

\section{Determination of $z$}
This appendix concerns the estimation of $\bar z$ from data collapses as
illustrated in Fig.~\ref{fig:scaling}. The problem we are faced with
is that we are looking for a collapse over a large range of
$x=tN^{-\bar z}$, and that the functional form of $Q(x)$ is not
easily expressed on a closed form or in low degree series
expansions. To get around this problem we partition the $x$-range in
$N_\mathrm{seg}$ segments $X_i$, $1\leq i\leq N_\mathrm{seg}$, and fit
a line ($a_i+b_ix$, $x\in X_i$) to the $Q$ point-set within each
segment (cf.~\cite{DFA}). Then we sum the square
of the deviations from the lines:
\begin{equation}
  \Lambda( {\bar z^\prime} ) =\sum_{0\leq i\leq N_\mathrm{seg}} \sum_{x\in X_i({\bar z^\prime})}
  (Q(x)-a_i-b_ix)^2~,
\end{equation}
where $X_i(\bar z^\prime)$ is the set of all numerical values of $tN^{-\bar z^\prime}$ and
thus depends on the  value of $\bar z^\prime$ chosen.
(Note that $t$ and hence $x$ are  discrete variables.) Now it is clear
that if the segmentation can be done so that $Q$ can be reasonably
well approximated by the line segments $a_i+b_ix$, i.e., if
$Q(x)$ is smooth enough, then ${\bar z}=\min_{\bar z^\prime}\Lambda(\bar z^\prime)$
will converge to the correct value as the number of samples and
$N_\mathrm{seg}$ are increased.

The remaining consideration is how to choose the segmentation. In
general, one needs the segments large enough to get a small error in
the linear regression, and small enough for the line-segment
approximation to be feasible. In practice the method seems to be
rather insensitive for the partition method. We choose to partition
the whole range of $x$ in segments of equal length, with
$N_\mathrm{seg} = 30$. The minimization of $\Lambda$ is conveniently
done by a Newton-Raphson method~\cite{NR}. The error in $z$ is
calculated by jackknife estimation~\cite{JACK}.


\begin{thebibliography}{99}

\bibitem{rev}
  S.~H.\ Strogatz, Science \textbf{284}, 79 (1999); S.~N.\
  Dorogovtsev, J.~F.~F.\ Mendes, Adv.\ Phys.\ \textbf{51}, 1079
  (2002); R.\ Albert and A.-L.\ Barab\'{a}si, Rev.\ Mod.\ Phys.\
  \textbf{74}, 47 (2002).
  
\bibitem{barr}
  A.\ Barrat and M.\ Weigt, Eur.\ Phys.\ J.\ B \textbf{13}, 547 (2000).
  
\bibitem{statswn} 
  B.~J.\ Kim, H.\ Hong, P.\ Holme, G.~S.\ Jeon, P.\ Minnhagen, and
  M.~Y.\ Choi, Phys.\ Rev.\ E \textbf{64}, 056135 (2001).

\bibitem{spin}
  M.~Gitterman, J.\ Phys.\ A: Math.\ Gen.\ \textbf{33}, 8373 (2000);
  P.~Svenson, Phys.\ Rev.\ E \textbf{64}, 036122 (2001);
  A.\ P\c{e}kalski, \textit{ibid.}\ \textbf{64}, 057104 (2001);
  H.~Hong, B.~J.\ Kim, and M.~Y.\ Choi, \textit{ibid.}\ \textbf{66},
  018101 (2002);
  C.~P.\ Herrero, \textit{ibid.}\ \textbf{65}, 066110 (2002);
  A.\ Aleksiejuk, J.~A.\ Holyst, and D.\ Stauffer, Physica A
  \textbf{310}, 260 (2002); 
  G.\ Bianconi, cond-mat/0204455;
  A.\ Aleksiejuk-Fronczak, e-print cond-mat/0206027; D.\ Boyer and O.\
  Miramontes, e-print cond-mat/0210352.
  
\bibitem{traffic}
  P.~Holme, e-print cond-mat/030113.

\bibitem{WS}
 D.~J.\ Watts and S.~H.\ Strogatz, Nature (London) \textbf{393}, 440 (1998). 

\bibitem{birds} J.\ Toner and Y.\ Tu, Phys.\ Rev.\ Lett.\ \textbf{75},
  4326 (1995).

\bibitem{shorttime}
  Z.~B.\ Li, L.\ Sch\"{u}lke, and B.\ Zheng, Phys.\ Rev.\ Lett.\ \textbf{74},
  3396 (1995). 

\bibitem{str} M.\ Silv\`{e}ro Soares, J.\ Kamphorst Leal da Silva, and
 F.~C.\ S\'{a} 
 Barreto, Phys.\ Rev.\ B \textbf{55}, 1021 (1997). 

\bibitem{luo}
  H.~J.\ Luo and B.\ Zheng, Mod.\ Phys.\ Lett.\ B \textbf{11}, 15412 (2000).

\bibitem{spinxy} B.~J. Kim, P. Minnhagen, S.~K. Oh, and J.~S. Chung,
Phys. Rev. B {\bf 64} 024406 (2001).

\bibitem{metropolis}
  N.\ Metropolis, A.~W.\ Rosenbluth, M.~N.\ Rosenbluth, A.~H.\ Teller, and
  E.\ Teller, J.\ Chem.\ Phys.\ \textbf{21}, 1087 (1953).

\bibitem{kim}
  B.~J.\ Kim, Phys.\ Rev.\ B \textbf{63}, 024503 (2001); K.~Medvedyeva,
  B.~J.\ Kim, and P.\ Minnhagen, Physica C \textbf{355}, 6 (2001).

\bibitem{scaling1} 
  L.~M.\ Jensen, B.~J.\ Kim, and P.\ Minnhagen, Phys.\ Rev.\ B \textbf{61},
  15412 (2000).

\bibitem{scaling2}
  S.\ Wansleben and D.~P.\ Landau, Phys.\ Rev.\ B \textbf{43}, 6006 (1991).

\bibitem{note}
  In $d$ dimensional regular lattices, we have $N=L^d$ and $\xi_V \sim
  \xi^{d}$, with 
  the linear size of the system $L$, the correlation length
  $\xi$, and the correlation volume $\xi_V$ . 
  Accordingly, at $T_c$ we have $\tau \sim L^{z} = N^{z/d} = N^{\bar z}$.
  Since $\xi$ diverges as $\xi \sim |T-T_c|^{-\nu}$, we also have
  $\xi_V \sim |T-T_c|^{-{\bar \nu}} = |T-T_c|^{-\nu d}$. 
  For systems where the correlation length is not well defined, but
  the correlation 
  volume is well-defined (as in the present work), 
  ${\bar z} = z/d$ and $\bar \nu = d \nu$ need to be measured to detect
  the dynamic and static universality class of the system. In the
  mean-field systems 
  corresponding to $d \geq 4$, $\bar z = 1/2$ and $\bar \nu = 2$ are
  expected.

\bibitem{newman}
  M.~E.~J.\ Newman and D.~J.\ Watts, Phys.\ Rev.\ E \textbf{60}, 7332 (1999).

\bibitem{ozana} 
  M.\ Barthelemy and L.~A.~N.\ Amaral, Phys.\ Rev.\ Lett.\ \textbf{82}, 3180
  (1999); \textbf{82}, 5180 (2000); M.\ Ozana, Europhys.\ Lett.\
  \textbf{55}, 762 (2001).

\bibitem{note1}
  If we assume a logarithmic correction to the characteristic length
  scale $\xi_V$, as it was suggested by A.~J.\ Bray,
  A.~J.\ Briant, and D.~K.\ Jervis, Phys.\ Rev.\ Lett.\ \textbf{84}, 1502
  (2000), we obtain a value $\bar z$ closer to 0.5.

\bibitem{note:z} To arrive at the result $\bar{z}=0.54(3)$ we set the
  threshold for breakdown of scaling (i.e.\ where we would have needed
  yet larger system sizes) to $P=0.06$, and average all points for
  larger $P$ values.

\bibitem{DFA} H.~E.\ Stanley, S.~V.\ Buldyrev, A.~L.\ Goldberger, S.\
  Havlin, C.-K.\ Peng, M.\ Simons, Physica A \textbf{200}, 4 (1993).

\bibitem{NR} W.~H.\ Press, S.~A.\ Teukolsky, W.~T.\ Vetterling, B.~P.\
  Flannery, \textit{Numerical recipes in C: The art of scientific computing}
  2nd ed.\ (Cambridge University Press, Cambridge, 1992).

\bibitem{JACK} B.\ Efron, \textit{The jackknife, the bootstrap, and other
  resampling plans} (SIAM, Philadelphia, 1982).

\end{thebibliography}
\end{document}